\documentclass[10pt,twocolumn,preprintnumbers,amsmath,amssymb,nofootinbib,superscriptaddress,aps,prd,numberedappendix]{openjournal}
\usepackage{newtxtext,newtxmath}
\usepackage{amsmath}
\usepackage{amssymb}
\usepackage{graphicx}
\usepackage{dcolumn}
\usepackage{bm}
\usepackage[dvipsnames]{xcolor}
\usepackage{xspace}
\usepackage{ulem}
\usepackage{lipsum}
\usepackage{xfrac}
\usepackage[colorlinks=true, allcolors=blue]{hyperref}
\newcommand{\nv}{\hat{\bf n}}
\newcommand{\tpm}{\dot{\boldsymbol{\alpha}}}
\newcommand{\Tpm}{\dot{\bf A}}
\newcommand{\Tpmcomp}{\dot{A}}
\newcommand{\uas}{\mu{\rm as}}
\newcommand{\uasy}{\mu{\rm as}\,{\rm yr}^{-1}}
\newcommand{\gaia}{{\sl Gaia}\xspace}

\begin{document}
\title{Can we constrain structure growth from galaxy proper motions?}

\author{Iain Duncan$^{1,*}$}
\author{David Alonso$^{1}$}
\author{An\v ze Slosar$^2$}
\author{Kate Storey-Fisher$^3$}
\email{$^*$iain.duncan@sjc.ox.ac.uk}
\affiliation{$^1$Department of Physics, University of Oxford, Denys Wilkinson Building, Keble Road, Oxford OX1 3RH, United Kingdom}
\affiliation{$^2$Brookhaven National Laboratory, Physics Department, Upton, NY 11973, USA}
\affiliation{$^3$Center for Cosmology and Particle Physics, Department of Physics, New York University, 726 Broadway, New York, NY 10003, USA}
\date{\today}

\begin{abstract}
  Galaxy peculiar velocities can be used to trace the growth of structure on cosmological scales. In the radial direction, peculiar velocities cause redshift space distortions, an established cosmological probe, and can be measured individually in the presence of an independent distance indicator. In the transverse direction, peculiar velocities cause proper motions. In this case, however, the proper motions are too small to detect on a galaxy-by-galaxy basis for any realistic experiment in the foreseeable future, but could be detected statistically in cross-correlation with other tracers of the density fluctuations. We forecast the sensitivity for a detection of transverse peculiar velocities through the cross-correlation of a proper motion survey, modelled after existing extragalactic samples measured by \gaia, and an overlaping galaxy survey. In particular, we consider a low-redshift galaxy sample, and a higher-redshift quasar sample. We find that, while the expected cosmological signal is below the expected statistical uncertainties from current data using cross-correlations, the sensitivity can improve fast with future experiments, and the threshold for detection may not be too far away in the future. Quantitatively, we find that the signal-to-noise ratio for detection is in the range $S/N\sim0.3$, with most of the signal concentrated at low redshifts $z\lesssim0.3$. If detected, this signal is sensitive to the product of the expansion and growth rates at late times, and thus would constitute an independent observable, sensitive to both background expansion and large-scale density fluctuations.
\end{abstract}
\maketitle

\section{Introduction}\label{sec:intro}
  Cosmology, perhaps more than any other physical science, suffers from lack of direct dimensional measurements. The majority of our data relies on angular positions, redshifts, and relative fluctuations. A small number of notable examples include the CMB temperature \citep{2009ApJ...707..916F}, direct measurements of the Hubble parameter through distance ladder \citep{2112.04510}, and strong lensing time delays \citep{2210.10833}.
  Recent advances in precision spectroscopy and astrometry can open new avenues for measuring cosmologically-induced changes in redshift and position of objects on the sky \citep{2203.05924}. Both measurements are related to time derivatives of cosmological quantities, and would thus be directly sensitive to the Universe's master clock rate, the Hubble parameter $H_0$. Since tensions in the measurements of the Hubble parameter are currently one of the main challenges for the standard $\Lambda$CDM cosmological model, exploring other observables sensitive to the expansion rate is of high priority.

  In this paper we investigate the possibility of detecting the transverse peculiar velocity field by measuring the correlated angular proper motions of galaxies, in particular considering their cross-correlation with other tracers of the cosmic density fluctuations, such as galaxy surveys. Physically, as the Universe evolves, the density contrast increases and matter flows towards over-dense regions. We therefore expect galaxy transverse peculiar velocities to preferably orient towards nearby over-densities. Statistically, this manifests itself as a non-zero correlation between the $E$-mode (i.e. the pure gradient component) of the proper motion field, and the galaxy overdensities. This constitutes the main observable we will consider here.

  As we will show, the corresponding signal is sensitive to the combination
  \begin{equation}
    H\,f\,\sigma_8,
  \end{equation}
  where $H$ is the Hubble parameter (i.e. the background expansion rate, with present value $H_0$), the {\sl growth rate} $f\equiv d\log\delta/d\log a$ is the logarithmic derivative of matter overdensities $\delta$ with respect to the scale factor $a$ in the linear regime, and $\sigma_8$ is the rms fluctuation of the density field in spheres of radius $R=8\,{\rm Mpc}/h$, with $h\equiv H_0/(100\,{\rm km}/{\rm s}/{\rm Mpc})$. Thus, besides offering us an independent measurement of the expansion rate, correlated proper motions are also sensitive to the growth and amplitude of matter perturbations. This addresses the so-called ``$S_8$ tension'' between measurements of the amplitude of inhomogeneities carried out by CMB and galaxy weak lensing experiments \citep{1809.09148,2105.13549,2105.13543,2105.13544,2204.02396,1502.01597,2105.03421,2206.10824,2007.15632,2105.12108}, another significant challenge for $\Lambda$CDM.

  A galaxy at $z\sim0.1$ moving across the sky at a typical transverse velocity of $300\,{\rm km/s}$ would cover an angle $\alpha\sim2\,\uas$ in 10 years of observation. With typical proper-motion uncertainties of at least $O(100)\,\uasy$ after a similar observation period, this signal is impossible to detect on a galaxy-by-galaxy basis. Measurements of the proper motion field caused by cosmological peculiar velocities will therefore be invariably noise-dominated. In terms of detectability, it is generally advantageous to correlate a noisy dataset with a high signal-to-noise tracer. For example, cross-correlations formed the basis of the first detection of CMB lensing \citep{0705.3980}, all existing measurements of the integrated Sachs-Wolfe effect \citep{2003ApJ...597L..89F,2010MNRAS.406....2F,2013MNRAS.430..259B,2015PhRvD..91h3533F,2022JCAP...04..033K,2016ApJ...827..116S}, and all evidence for the clustering of damped Lyman-$\alpha$ systems \citep{1209.4596,1709.00889,1712.02738}. While detecting transverse galaxy peculiar motions has long been considered to be unfeasible, we will show that \gaia \citep{2016A&A...595A...1G}, the European Space Agency's flagship astrometric mission, may come close in raw sensitivity to the detection threshold when combined with a sufficiently dense galaxy survey at low redshifts (where amplitude of the proper motion signal peaks). This could thus be transformed into a high-significance detection with a future, more sensitive mission, unlocking a new probe of structure growth in the low-redshift Universe.

  Previous literature has studied different approaches to cosmological proper motions. \cite{1811.05454} has focused on the secular parallax due to the Solar System's motion relative to distant objects \citep[see also][]{2013ApJ...777L..21D}. \cite{1807.06658} have investigated the configuration-space correlation of galaxy proper motions and forecasted the sensitivity of the auto-correlation measurements with results that are broadly consistent with ours for that observable.

  This paper is structured as follows. In Section \ref{sec:theory} we develop the mathematical background needed to quantify the expected cosmological signal from peculiar motion measurements, its statistical uncertainties, and its detectability for different experimental configurations. Section \ref{sec:data} briefly discusses the model used to describe a generic proper motion experiment based on the current performance of \gaia. Our results concerning the detectability of this signal are presented in Section \ref{sec:res}, and summarised in Section \ref{sec:conc}.

\section{Theory}\label{sec:theory}
  \subsection{Proper motions and peculiar velocities}\label{ssec:theory.tpm}
    Let $\nv$ be the unit vector pointing in the direction of a given source. The observable we aim to study here is the measured transverse peculiar motion, defined as
    \begin{equation}
      \tpm\equiv\frac{d\nv}{dt_o},  
    \end{equation}
    where $t_o$ is the time measured at the observer's position. Our theoretical description for this observable follows and builds upon the derivation of \cite{1811.05454}, focusing on the projected (angular) statistics of the transverse velocity field, and deriving a prediction for its angular power spectrum and cross-correlation with galaxies. Ignoring general-relativistic corrections -- since they are generally small \citep{1507.03550}, and the comoving distances we are most sensitive to are smaller than the horizon -- time intervals at the observer and the emitter are related via $dt_o=(1+z)\,dt$, where $t$ is the time measured at the emitter's rest frame. Writing $\nv$ in terms of the source's comoving coordinates ${\bf x}$ as $\nv\equiv{\bf x}/\chi$, where $\chi$ is the radial comoving distance, we can relate $\tpm$ to the transverse comoving velocity ${\bf v}_\perp\equiv(1-\nv^T\nv)\cdot{\bf v}=\dot{\bf x}-\nv\dot{\chi}$, to obtain the intuitive result
    \begin{equation}\label{eq:alpha_to_vt}
      \tpm=\frac{{\bf v}_\perp}{(1+z)\chi}.
    \end{equation}
    
    After measuring $\tpm$ at the position of several objects, we can now make maps of the projected peculiar motion, by averaging over the values of $\tpm$ for all galaxies lying on each pixel. Let us call the resulting map $\Tpm$, and let $p(z)$ be the distribution of source redshifts. The value of $\Tpm$ measured along the pixel centered around position $\nv$ is then
    \begin{equation}
      \Tpm(\nv)=\int dz\,p(z)\,\tpm(\chi\nv).
    \end{equation}

    Neglecting vorticity modes (which are heavily suppressed on large scales), the comoving peculiar velocity field can be written in terms of velocity potential ${\bf v}=\nabla \varphi_v$. $\Tpm$ can therefore be expressed as the angular gradient\footnote{The angular gradient $\nabla_\theta$ is related to the transverse gradient $\nabla_\perp$ via $\nabla_\perp=\chi^{-1}\nabla_\theta$.} of an effective projected potential $\Phi_v$, given by
    \begin{equation}\label{eq:Phi}
      \Phi_v(\nv)=\int dz\,\frac{p(z)}{\chi^2(1+z)}\,\varphi_v(\chi\nv).
    \end{equation}
    The two components of the projected velocity map can then be combined into a complex spin-1 field $\Tpmcomp\equiv \Tpmcomp_1+i\Tpmcomp_2$, which is related to $\Phi_v$ through the differential spin-raising operator $\eth$ \citep{Goldberg_1967} as
    \begin{equation}\label{eq:eth}
      \Tpmcomp=-\eth\Phi_v.
    \end{equation}
    Since $\Tpmcomp$ is a pure gradient, it is also a pure $E$-mode field (i.e. its $B$-mode component should be zero). In practice, a $B$-mode component would be generated by any vortical contribution to ${\bf v}$ (e.g. due to non-linear gravitational collapse), or by source clustering effects in the $\Tpm$ map (for weak gravitational effect lensing equivalent see \cite{astro-ph/0112441}), both of which are higher-order effects in perturbation theory, and thus negligible given the statistical uncertainties of this observable (see Appendix \ref{app:lensing}). The spherical harmonic coefficients of the $E$-mode component of $\Tpmcomp$ are then simply related to those of $\Phi_v$ via
    \begin{equation}\label{eq:ellellp1}
      \Tpmcomp^E_{\ell m}=\sqrt{\ell(\ell+1)}\Phi_{v,\ell m}.
    \end{equation}

    In order to relate the peculiar velocity potential to the matter overdensity $\delta$, we make use of the linear continuity equation ($\dot{\delta}+\nabla\cdot{\bf v}=0$), which in Fourier space reads
    \begin{equation}
      \varphi_v({\bf k})=\frac{Hf}{k^2}\delta({\bf k}).
    \end{equation}
    Here $f\equiv d\log\delta/d\log a$ is the growth rate of structure, with $a$ the scale factor, $H\equiv\dot{a}/a$ is the expansion rate, and we have assumed self-similar growth (valid in the linear regime in $\Lambda$CDM). Thus we see that the peculiar velocity field is sensitive to the combination $H\,f$, as well as the amplitude of the matter overdensity $\delta$, commonly parametrised in terms of $\sigma_8$ (the standard deviation of $\delta$ on spheres of comoving radius $R=8\,h^{-1}{\rm Mpc}$). As we will see, the scales over which the transverse peculiar motion field can be detected are rather large, and the linear approximation above should be sufficiently accurate. We leave a more thorough study of the impact of various non-linear effects (including the generation of $B$-modes) for future work.

    Before we move on to calculate the cross-correlation between proper motion maps and the galaxy overdensity, it is worth noting that a variation in the angular positions of distant sources may also be caused via gravitational lensing by time-varying gravitational potentials. As we show in Appendix \ref{app:lensing}, however, the signal is only sourced at low redshifts and is always at least one order of magnitude lower than the peculiar velocity contribution.

  \subsection{Galaxy overdensity}\label{ssec:theory.gals}
    The second observable we will use in this work is the projected galaxy overdensity $\Delta_g$, given in terms of the 3D overdensity $\delta_g$ as
    \begin{equation}\label{eq:deltag}
      \Delta_g(\nv)\equiv\int dz\,p(z)\,\delta_g(\chi\nv),
    \end{equation}
    where $p(z)$ is the redshift distribution of the sample under study. We will use a simple linear bias relation to relate the galaxy and matter overdensities:
    \begin{equation}
      \delta_g=b_g\,\delta.
    \end{equation}
    On the large scales over which the peculiar motion signal can be detected this should be sufficiently accurate \citep{2307.03226}.

  \subsection{Angular power spectra}\label{ssec:theory.cl}
    Consider now a projected velocity map constructed from a galaxy sample with redshift distribution $p_A(z)$, and a separate galaxy sample with distribution $p_g(z)$, from which we construct a map of the projected overdensity field. The angular power spectrum $C_\ell$ is defined to be the co-variance between the harmonic coefficients of any pair of such maps:
    \begin{equation}
      \langle a_{\ell m}b_{\ell'm'}\rangle\equiv C_\ell^{ab}\,\delta_{\ell\ell'}\delta_{mm'}.
    \end{equation}

    Using the results from the two preceding sections, we can relate the power spectra of the projected galaxy overdensity and the peculiar velocity $E$-mode to the matter power spectrum $P(k)$ via 
    \begin{equation}\label{eq:cl_full}
      C_\ell^{ab}=\frac{2}{\pi}\int dk\,k^2\left[\prod_{i\in\{a,b\}}\int d\chi_i\,Q^i_\ell(\chi_i,k)\right]\,P(k,z),
    \end{equation}
    where we have defined the transfer functions for the galaxy overdensity ($Q^g$) and for the proper motion map $E$-mode ($Q^A$):
    \begin{align}
      &Q^g_\ell(k,\chi)\equiv \frac{dz}{d\chi}\,p_g(z)b_g(z)\,j_\ell(k\chi),\\
      &Q^A_\ell(k,\chi)=\frac{dz}{d\chi}\,\frac{\sqrt{\ell(\ell+1)}}{(k\chi)^2}\,p_A(z)\,\frac{H(z)\,f(z)}{1+z}j_\ell(k\chi).
    \end{align}
    Here, $dz/d\chi=H(z)$, and $j_\ell(k\chi)$ is the order-$\ell$ spherical Bessel function\footnote{Note that we use natural units with the speed of light $c=1$.}. The functional form of the transfer functions is determined by the projection kernels associated with the two quantities under study (see Eq. \ref{eq:Phi} and Eq. \ref{eq:deltag}). The factor $\sqrt{\ell(\ell+1)}$ originates from the angular derivative in Eq. \ref{eq:eth} (see Eq. \ref{eq:ellellp1}).
    
    If the kernels under study are sufficiently broad, the computation of Eq. \ref{eq:cl_full} can be accelerated through the use of Limber's approximation \citep{1953ApJ...117..134L,1992ApJ...388..272K}. Note, however, that we will not make use of this approximation in any of the calculations presented here. Since galaxy peculiar velocities are uncorrelated at large distances, averaging the proper motion of galaxies over large radial separations would wash out the signal, and thus this probe benefits from using narrow redshift bins, for which Limber's approximation is not appropriate. This also implies that a practical application of this probe would benefit from a fully three-dimensional treatment, where the transverse velocity and galaxy overdensity fields are reconstructed and correlated in 3D. We leave the development of this formalism (which would follow closely the 3D treatment of intrinsic alignments described in \cite{2009.00276}) for future work.

  \subsection{Noise model}\label{ssec:theory.noise}
    To forecast the detectability of the cosmic transverse velocity signal we need to quantify the noise properties of the projected proper motion and galaxy overdensity maps.
    
    The only source of noise for $\Delta_g$ is Poisson noise due to the discrete number of sources. The corresponding noise power spectrum is
    \begin{equation}
      N_\ell^{gg}=\frac{1}{\bar{n}_g},
    \end{equation}
    where $\bar{n}_g$ is the angular number density of galaxies in the sample, in units of ${\rm srad}^{-1}$.
    
    The peculiar motion map is constructed by averaging, in each pixel, the peculiar motion vector $\tpm$ of each galaxy in the pixel (the map is therefore, in fact, two maps, one for each component). The noise power spectrum of the $E$-mode of the resulting map is then given by
    \begin{equation}
      N^{AA}_\ell=\frac{\sigma_{\rm pm}^2}{\bar{n}_A},
    \end{equation}
    where $\bar{n}_A$ is the number density of sources with measured peculiar motions, and $\sigma_{\rm pm}$ is the root mean square error in the measurement of each component of $\tpm$ for a given galaxy.

    To estimate $\sigma_{\rm pm}$, consider observations of the position ${\bf x}$ of a given source at various times: $\{{\bf x}_n\equiv{\bf x}(t_n),\,\,n\in[1,N]\}$, where $N$ is the total number of visits. The apparent peculiar motion of this object, $\tpm$, can be measured by modelling it as:
    \begin{equation}
      x_{i,n}=\dot{\alpha}_it_n+n_{i,n},
    \end{equation}
    where the index $i$ runs over the two sky coordinates, and $n_{i,n}$ is random noise caused by astrometric uncertainties. The value of $\dot{\alpha}_i$ can be estimated from these measurements using linear regression, under the assumption that the noise $n_{i,n}$ is Gaussianly distributed with standard deviation $\sigma_{x,n}$. In this case, the per-component uncertainty on $\tpm$ is given by
    \begin{equation}
      \sigma^2_{\rm pm}=\left[\sum_{n=1}^N\frac{t_n^2}{\sigma_{x,n}^2}\right]^{-1}.
    \end{equation}
    For simplicity, let us assume periodic visits, so $t_n=nT/N$, where $T$ is the total observing time, and that the astrometry uncertainty is constant $\sigma_{x,n}=\sigma_x$. The proper motion variance then reads
    \begin{equation}
      \sigma^2_{\rm pm}=\frac{\sigma_x^2N^2}{T^2}\frac{1}{\sum_nn^2}.
    \end{equation}
    Using $\sum_{n=1}^Nn^2=N(N+1)(2N+1)/6\simeq N^3/3$ in the large-$N$ limit, we obtain:
    \begin{equation}\label{eq:sigma_alpha}
      \sigma_{\rm pm}=\sqrt{3}\frac{\sigma_x}{T\sqrt{N}}.
    \end{equation}
    Assuming a constant cadence ($N\propto T$), the peculiar motion uncertainties thus decrease with survey time as
    \begin{equation}
      \sigma_{\rm pm}\propto T^{-3/2}.
    \end{equation}
    This is significantly faster than the usual $\propto\sqrt{T}$ scaling of mapping experiments. It implies that an order of magnitude improvement in the map-level noise rms can be achieved by reducing astrometry errors by a similar factor, or by increasing the total observation time only by a factor $\sim4.6$. 

  \subsection{Forecasting detectability}\label{ssec:theory.sn}
    We quantify the significance with which the peculiar motion signal can be detected as the square root of the Fisher matrix element for an effective amplitude parameter, with fiducial value $1$, multiplying the signal component of the peculiar motion map $\Tpmcomp(\nv)$. For the cross-correlation between $\Tpmcomp$ and a galaxy overdensity map, the signal-to-noise ratio is
    \begin{equation}\label{eq:sn_x}
      \left(\frac{S}{N}\right)^2_{gA}=\sum_{\ell=\ell_{\rm min}}^{\ell_{\rm max}}\frac{(2\ell+1)\left(C_\ell^{gA}\right)^2}{(C_\ell^{gg}+N_\ell^{gg})(C_\ell^{AA}+N_\ell^{AA})+\left(C_\ell^{gA}\right)^2}.
    \end{equation}
    In the case of the peculiar motion auto-correlation, in turn
    \begin{equation}
      \left(\frac{S}{N}\right)^2_{AA}=\sum_{\ell=\ell_{\rm min}}^{\ell_{\rm max}}=\frac{2(2\ell+1)\left(C_\ell^{AA}\right)^2}{\left(C_\ell^{AA}+N_\ell^{AA}\right)^2}.
    \end{equation}

    The cross-correlation is therefore likely to achieve significantly better $S/N$ ratio than the auto-correlation if the peculiar motion map is noise-dominated. This is because, in this case, the cross-correlation $S/N$ scales as $(N_\ell^{AA})^{-1/2}\propto\sigma_{\rm pm}^{-1}\propto T^{3/2}$, instead of $(N_\ell^{AA})^{-1}\propto\sigma_{\rm pm}^{-2}\propto T^3$ for the auto-correlation.

\section{Proper motion surveys}\label{sec:data}
  \begin{figure*}
    \centering
    \includegraphics[width=0.49\textwidth]{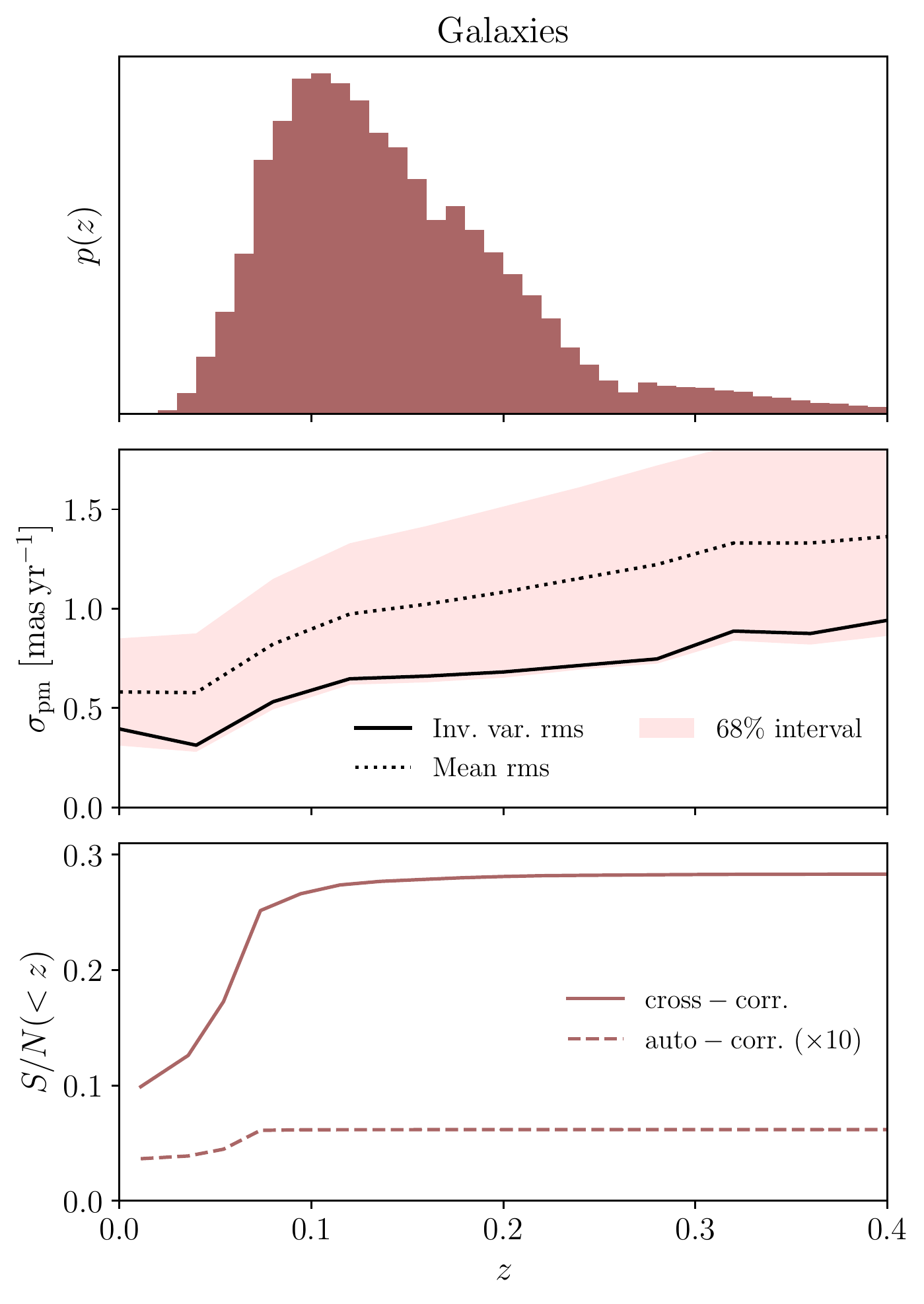}
    \includegraphics[width=0.49\textwidth]{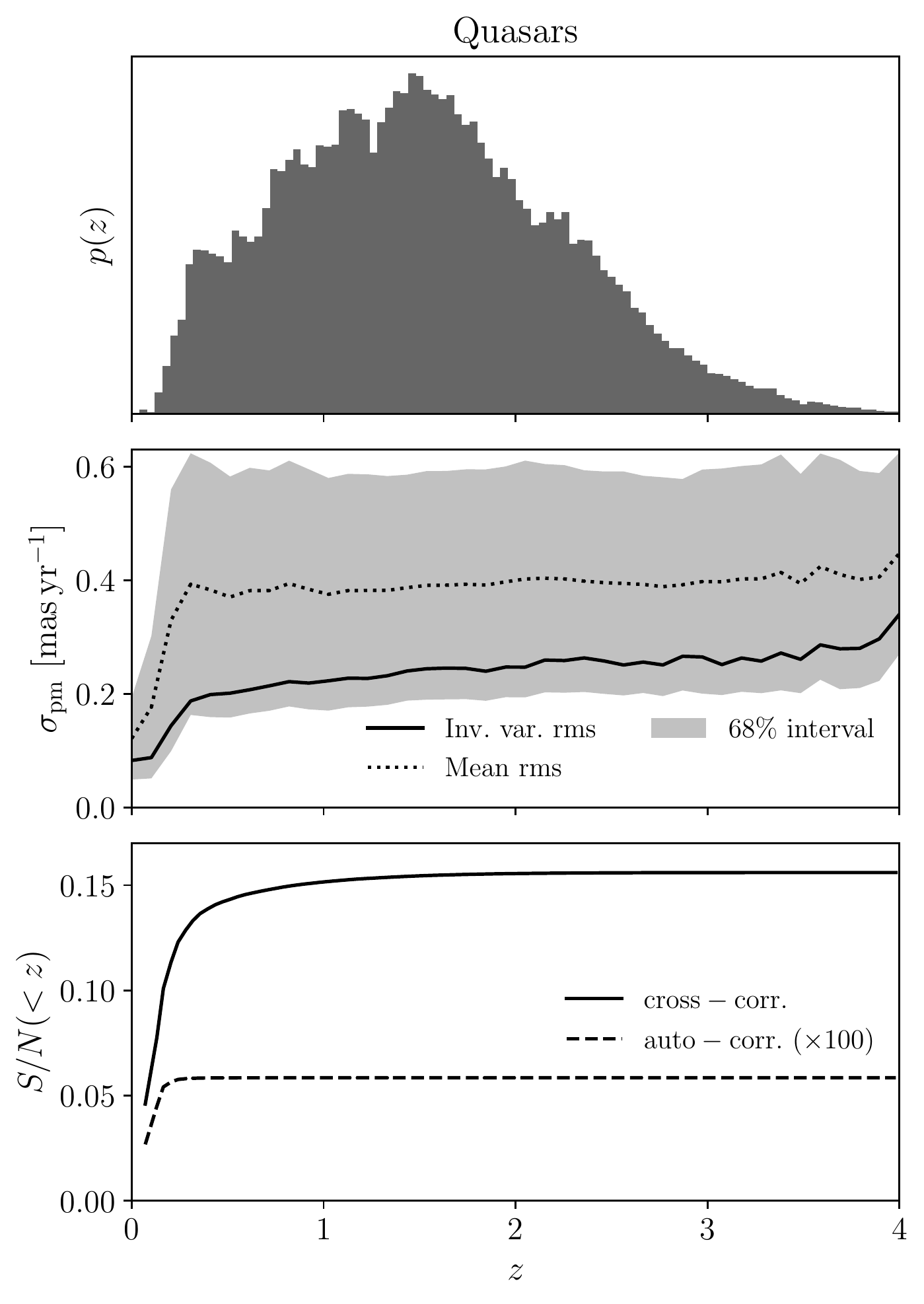}
    \caption{Redshift distribution (top row), proper motion rms error (middle row), and cumulative signal-to-noise ratio (bottom row) as a function of redshift for the \gaia galaxy and quasar samples (left and right columns respectively). In the middle pannels, the dotted lines show the mean rms proper motion error $\sigma_{\rm pm}$ for all sources at a given redshift, with the shaded band showing the standard deviation of this quantity. The solid lines show the inverse-variance-weighted mean rms error (which we use to calculate the map-level noise of the proper motion maps). The $S/N$ is shown for the cross-correlation with a galaxy sample (solid line), and for the proper motion autocorrelation (dashed line). The auto-correlation result is multiplied by 10 and 100 for galaxies and quasars respectively, so it can be shown on the same scale. The cross-correlation result assumes that the proper motion map is correlated with a galaxy sample having sufficiently high density that shot noise can be ignored.}
    \label{fig:samples}
  \end{figure*}
  To model a generic astrometric survey that could potentially measure the cosmic proper motion signal, we use data from the \gaia DR3 data release\footnote{\url{https://www.cosmos.esa.int/web/gaia/dr3}} \citep{2208.00211}. This data release corresponds to about 34 months of data collection. We focus on the two main extragalactic source types made available with the release, galaxies and quasars; the parent samples are detailed in \cite{gaiacollaboration2022gaia}. These sources have proper motion and redshift information, and cover the entire sky (besides the galactic plane). Our main use of this data is to model the distribution of errors in our forecasting exercise. The specific samples we use are:
  \begin{itemize}
    \item {\bf Low-redshift galaxies:} \gaia DR3 contains a sample of $\sim1.4\times10^6$ galaxies at $z\lesssim0.4$, of which $\sim3.7\times10^5$ have proper motion measurements and redshift estimates. Given the extended nature and lower surface brightness of galaxies, compared to the quasar sample, the proper motion uncertainties of this sample are significantly higher, ranging from $\sim500$ to $\sim1400\,\uasy$. However, the angle covered by a galaxy moving at a fixed transverse velocity scales as $\chi^{-1}$ with radial distance, and thus the amplitude of the proper motion signal grows quickly towards $z=0$, compensating for the comparably larger measurement noise.
    \item {\bf High-redshift quasars:} \gaia DR3 contains millions of quasar candidates. This parent sample has very low purity, containing many stellar contaminants, so we use the clean \gaia-unWISE quasar catalog derived from it by Storey-Fisher et al. (in prep.). This contains $\sim1.3 \times 10^6$ quasars with \gaia $G$-band magnitude less than 20.5, and also has improved redshift estimates which span the redshift range $z\lesssim4$. Given the almost point-source nature of quasars, they constitute an ideal extragalactic sample for proper motion measurements. Typical proper motion uncertainties for this sample range from $\sim150$ to $\sim350\,\uasy$. A primary use for quasar proper motions is to calibrate the motion of Milky Way stars in the same sky patch. Instead, here we will use this sample as a potential tracer of the transverse velocity field. The calibration of the quasar proper motions themselves is a challenging problem that we will not explore here.
  \end{itemize}

  Fig. \ref{fig:samples} shows the redshift distribution and typical proper motion uncertainties of both samples. It is worth noting proper motion uncertainties vary significantly from one object to another (the shaded bands in the figure show the standard deviation of $\sigma_{\rm pm}$). Thus, our estimate of the map-level noise in each redshift bin will assume that the proper motion map has been constructed by averaging over sources using inverse-variance weighting:
  \begin{equation}
    \bar{\sigma}_{\rm pm}^{-2}=\frac{1}{N}\sum_{i=1}^N\sigma_{{\rm pm},i}^{-2},
  \end{equation}
  where the index $i$ runs over the $N$ sources in the redshift bin. This estimate of the map-level noise is shown as a solid line in the figure, in contrast with the mean rms error, shown as a dotted line.
  
  These parameters (number of galaxies, distribution in redshift, and proper motion errors) will characterise the fiducial setup explored in the next section. In all cases we will also explore improved versions of this fiducial (e.g. with a larger number of sources, lower proper motion errors, or longer observation times).

\section{Results}\label{sec:res}
  We combine the methods described above to forecast the signal-to-noise of different samples. We will treat the galaxy and quasar catalogs of \gaia DR3 as our fiducial setup, defined by the redshift distribution and proper motion errors shown in the two top rows of Fig. \ref{fig:samples}, and the sample sizes quoted in Section \ref{sec:data}. Unless otherwise stated, we will consider only the cross-correlation between proper motion maps and a sample of galaxies. The latter will, by default, be assumed to be unaffected by shot noise (i.e. we will report results in the limit $\bar{n}_g\rightarrow\infty$), and we will explore the impact of shot noise in Section \ref{ssec:res:shot}.
  
  To produce our forecasts we will divide the \gaia DR3 galaxy and quasar samples into top-hat redshift bins with constant width $\delta z=0.04$. This leads to 10 bins in the range $z<0.4$ for the galaxy sample, and 100 bins in the range $z<4$ for the quasar sample. We note that the forecast of the signal-to-noise ratio $S/N$ is somewhat sensitive to this choice, emphasizing that a practical analysis of proper motion cross-correlations should be carried out using a fully three-dimensional formalism. Nevertheless, we find that the results presented here change by less than 10$\%$ when doubling the number of bins. We deem this level of precision to be accurate enough for our forecasts, given the large uncertainties in other aspects of this analysis (e.g. the impact of large-scale systematics in the proper motion maps).

  By default, in all cases we will report forecast signal-to-noise ratios using only multipoles $\ell>2$. Note that the peculiar motion signal should have a non-zero dipole ($\ell=1$). Reliably measuring this dipole, is likely challenging \citep{1811.05454}, and hence we focus on the higher multipoles. We will quantify the improvement in expected $S/N$ when including the $\ell=1$ mode in Section \ref{ssec:res:ell}.

  \subsection{Forecasts for the fiducial \gaia samples}\label{ssec:res.fid}
    The lower panels of Fig. \ref{fig:samples} show the cumulative signal to noise ratio for each of the two samples as a function of the maximum redshift included in the analysis. Results are shown for the cross-correlation with a dense low-redshift sample as solid lines, and for the proper motion auto-correlation as dashed lines. Note the multiplicative factors of 10 and 100 applied to the auto-correlation result in order to show both lines on the same scale.
    
    Several important conclusions can be derived from this figure. First, the signal is dominated by sources at low redshifts ($z\lesssim0.1$ for the galaxy sample, and $z\lesssim0.5$ for the quasar sample). This is to be expected, given the inverse scaling of proper motions induced by peculiar velocities with radial distance (Eq. \ref{eq:alpha_to_vt}). Secondly, in spite of its significantly larger measurement uncertainties $\sigma_{\rm pm}$, the galaxy sample achieves a $S/N$ that is twice that of the quasar sample. This is because the larger measurement noise is compensated by the significantly higher number density concentrated at low redshifts for the galaxy sample. Thirdly, the signal-to-noise of the auto-correlation is orders of magnitude smaller than the cross-correlation in all cases. As we discussed in Section \ref{ssec:theory.sn}, this is to be expected given the milder scaling with measurement error of the cross-correlation in the noise-dominated regime.

    Finally, integrating over the full redshift range, we conclude that the current \gaia samples should be able to ``detect'' the cosmological proper motion signal (as predicted by $\Lambda$CDM) with a signal-to-noise ratio
    \begin{equation}
      \left(\frac{S}{N}\right)_{\rm gal.}=0.3,\hspace{12pt}\left(\frac{S}{N}\right)_{\rm QSO}=0.15,
    \end{equation}
    for the galaxy and quasar sample respectively. The current data sets are not sensitive enough for a convincing detection ($S/N\sim 5$) of this proper motion signal. However, the sensitivity that is forecast to achieve this goal is only a factor $\sim10$ (as opposed to several orders of magnitude) smaller than the rough detection limit, and thus it is not far-fetched to imagine a future experiment improving on the sensitivity of \gaia that could make such measurement.

  \subsection{Forecasts for future surveys}\label{ssec:res.future}
    \begin{figure}
      \centering
      \includegraphics[width=0.49\textwidth]{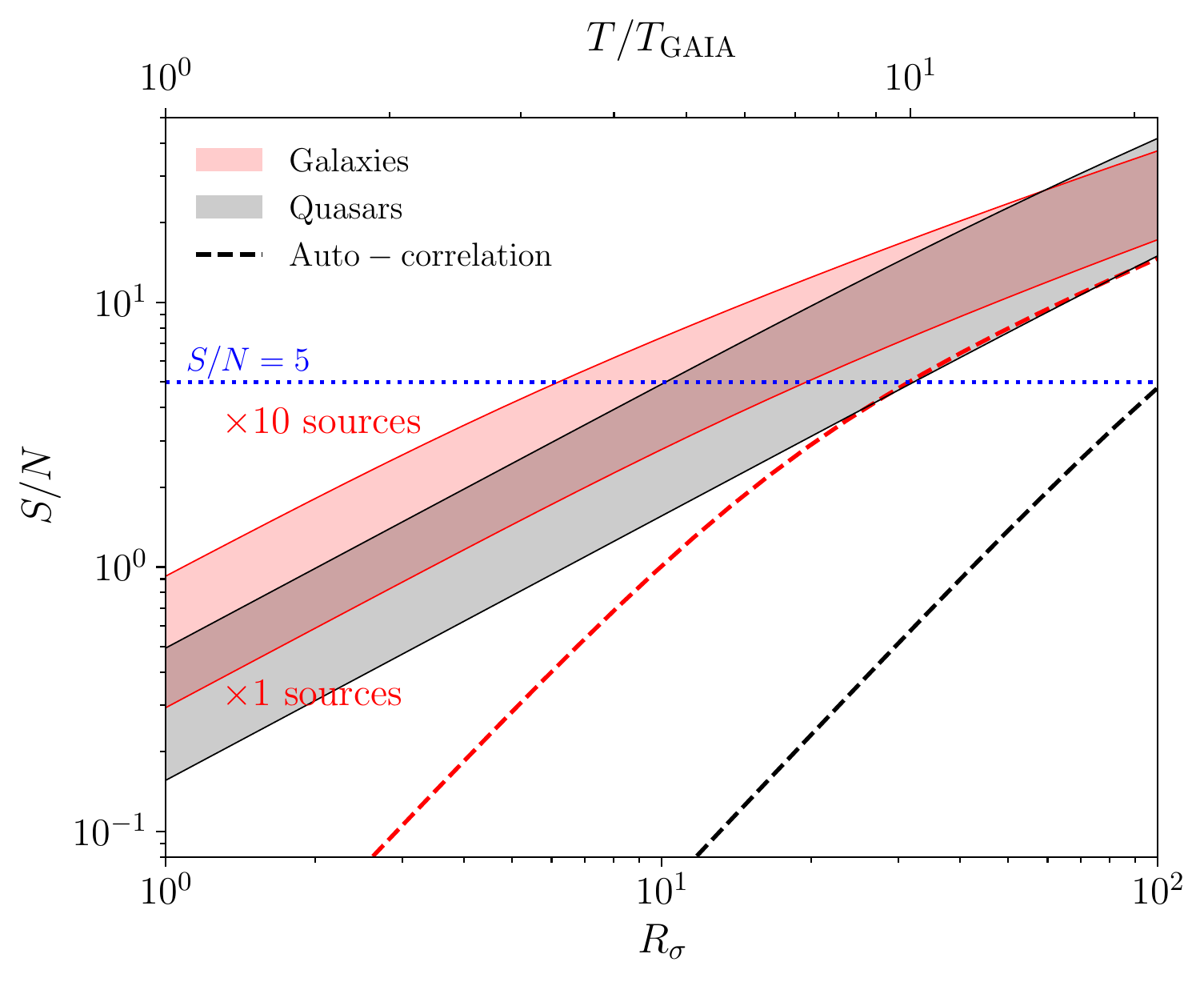}
      \caption{Signal-to-noise ratio of the proper motion signal measured via cross-correlations (shaded bands) and the auto-correlation (dashed lines) as a function of the factor by which the current sensitivity achieved by \gaia would need to be improved (see main text for a precise definition). The shaded bands show the range of $S/N$ achieved for samples containing a factor $R_n$ more sources than the current \gaia catalogs, with $1<R_n<10$. Results are shown in red and black for the \gaia galaxy and quasar samples, respectively. The upper $x$ axis shows the equivalent factor by which the observing time of \gaia would need to be increased to achieve a similar precision.}
    \label{fig:SN_sens}
    \end{figure}
    Since existing datasets may not be that far from detecting the cosmological proper motion signal, the aim of this section is to forecast the improvement in sensitivity that would be needed in order to make this detection with a future experiment. As we discussed in Section \ref{ssec:theory.noise}, the proper motion noise can be reduced by reducing the astrometric uncertainties ($N_\ell^{AA}\propto\sigma_{\rm pm}^2\propto\sigma_x^2$), increasing the observation time ($N_\ell^{AA}\propto\sigma_{\rm pm}^2\propto T^{-3}$), or increasing the number of sources with measured proper motions ($N_\ell^{AA}\propto \bar{n}_A^{-1}$). 

    Figure \ref{fig:SN_sens} shows the signal-to-noise ratio of a hypothetical experiment as a function of the factor $R_\sigma$ by which proper motion uncertainties have improved with respect to \gaia. I.e.
    \begin{equation}
      R_\sigma\equiv \frac{\sigma_{{\rm pm},Gaia}}{\sigma_{\rm pm}}.
    \end{equation}
    We consider samples based on the \gaia galaxy and quasar catalogs, shown in red and black respectively. The two shaded bands show the range of $S/N$ that would be achieved by increasing the number of sources in the samples by a factor $R_n$, with $R_n=1$ (i.e. current size) corresponding to the lower bound, and $R_n=10$ corresponding to the upper bound. The upper $x$-axis shows the factor by which the observation time of \gaia would need to be extended, keeping astrometric errors constant, to achieve a given $R_\sigma$.
    
    We find that current samples could achieve $S/N\sim1$ by either increasing the sample size by a factor $\sim10$, or improving proper motion measurement sensitivities by a factor $R_\sigma\sim4$ (corresponding to a factor $\sim2.5$ increase in observation time with respect to \gaia). Reaching $S/N=5$ (marked by a horizontal blue dotted line in the figure) may require a factor $\sim6$ increase in observation time ($R_\sigma\sim15$), or a $\sim3$ times larger observation time accompanied by a 10-fold increase in sample size. The sensitivity of the quasar sample remains lower than that of the galaxy sample, except in the high-sensitivity regime ($R_\sigma\sim70$, $R_n=10$), where the galaxy sample becomes limited by cosmic variance.

    The dashed lines in the same figure show the corresponding $S/N$ achieved with the auto-correlation. In this case the $S/N$ curves increase more steeply with $R_\sigma$ ($(S/N)_{\rm auto.}\propto R_\sigma^2$, whereas $(S/N)_{\rm cross.}\propto R_\sigma$ in the noise-limited regime), and $S/N\sim1$ can be achieved at $R_\sigma\sim10$ (or a 5-fold increase in observation time) with the auto-correlation. In the cosmic-variance-limited regime ($R_\sigma\gtrsim100$), the auto- and cross-correlations achieve comparable sensitivities.

  \subsection{Sensitivity as a function of scale}\label{ssec:res:ell}
    \begin{figure}
      \centering
      \includegraphics[width=0.49\textwidth]{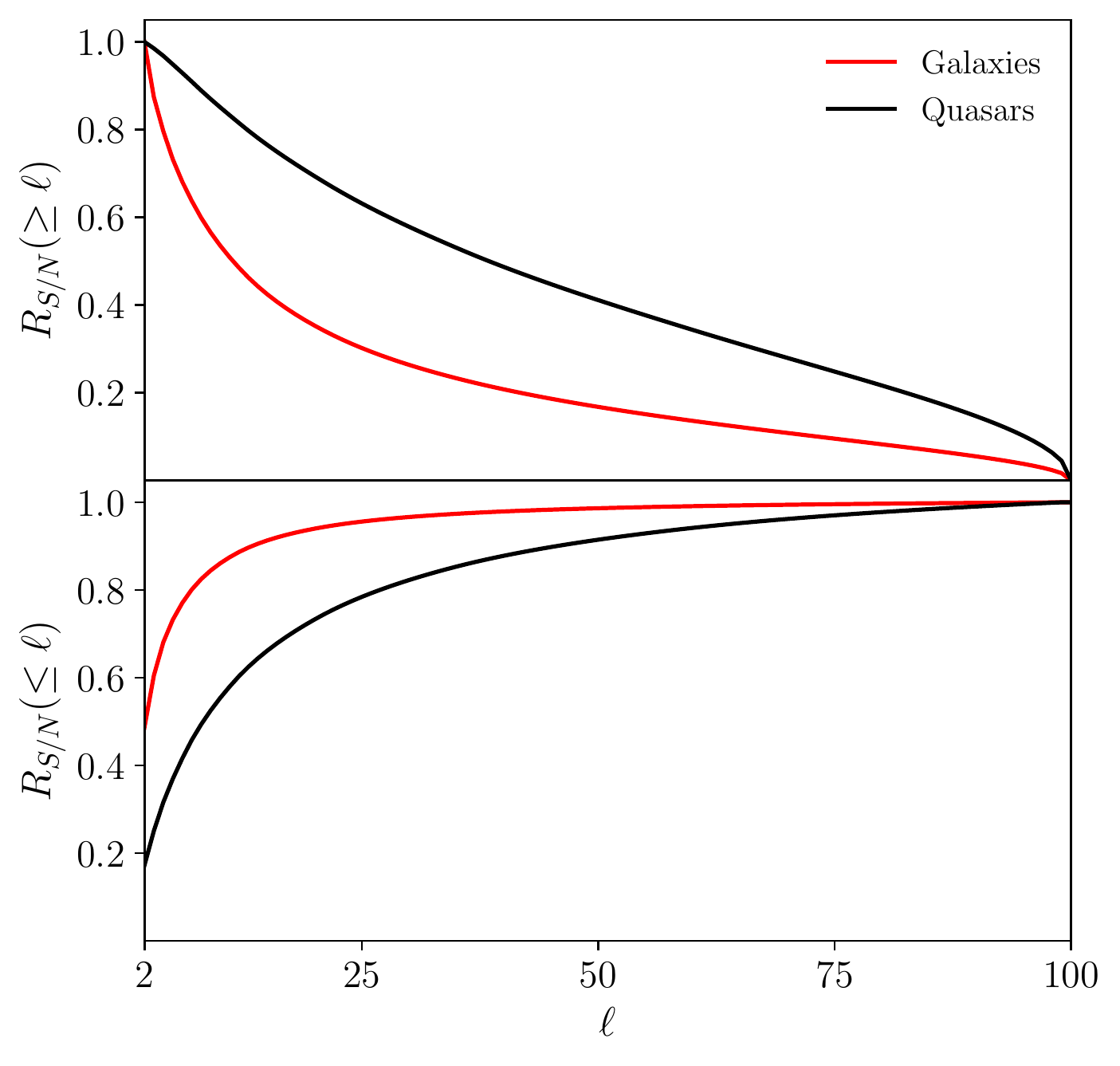}
      \caption{Fraction of the total signal-to-noise ratio contained in multipoles larger (upper panel) or smaller (lower panel) than a given $\ell$. Results are shown for the \gaia galaxy and quasar samples in red and black respectively.}
    \label{fig:SN_ells}
    \end{figure}
    Since measurements of the proper motion signal on large scales are likely affected by various systematics, it is worth quantifying the detectability of this signal as a function of the range of angular scales included in the analysis. Figure \ref{fig:SN_ells} shows the fraction of the total signal to noise (labelled $R_{S/N}$) as a function of the minimum and maximum multipoles used (top and bottom panels respectively). For the \gaia galaxy sample, shown in red, we find that $\sim90\%$ of the signal is concentrated on scales $\ell\lesssim25$, and that a measurement of the quadrupole ($\ell=2$) would achieve approximately half of the total signal-to-noise ratio. The quasar sample, on the other hand, has a milder dependence on scale, and $\sim50\%$ of the sensitivity can still be recovered with scales $\ell\gtrsim50$. This is due to the higher effective redshift of the sample, which pushes part of the signal to larger multipoles.

    As we discussed above, we have so far ignored the information encoded in the dipole ($\ell=1$). However, given the steep scale dependence observed in Fig. \ref{fig:SN_ells}, we can expect the dipole to have a non-negligible constribution to the total sensitivity. We find that including the $\ell=1$ mode leads to improved sensitivities by a factor
    \begin{equation}
     \left.\frac{(S/N)_{\ell\geq1}}{(S/N)_{\ell\geq2}}\right|_{\rm gal.}=1.59,
     \hspace{12pt}
     \left.\frac{(S/N)_{\ell\geq1}}{(S/N)_{\ell\geq2}}\right|_{\rm QSO}=1.01.
    \end{equation}
    The dipole is therefore particularly important for the low-redshift galaxy sample, potentially even dominating the total signal-to-noise ratio. It would be therefore imperative in any attempt to measure the cosmic proper motion signal from low-redshift samples, to conduct a thorough study of the dominant systematics affecting proper motion measurements on the largest angular scales. For a higher-redshift sample, in turn, the dipole contribution is significantly smaller, and such a sample would rely less on the precision with which large-scale systematics in proper motion measurements can be kept under control.

  \subsection{The impact of shot noise and bias}\label{ssec:res:shot}
    \begin{figure}
      \centering
      \includegraphics[width=0.49\textwidth]{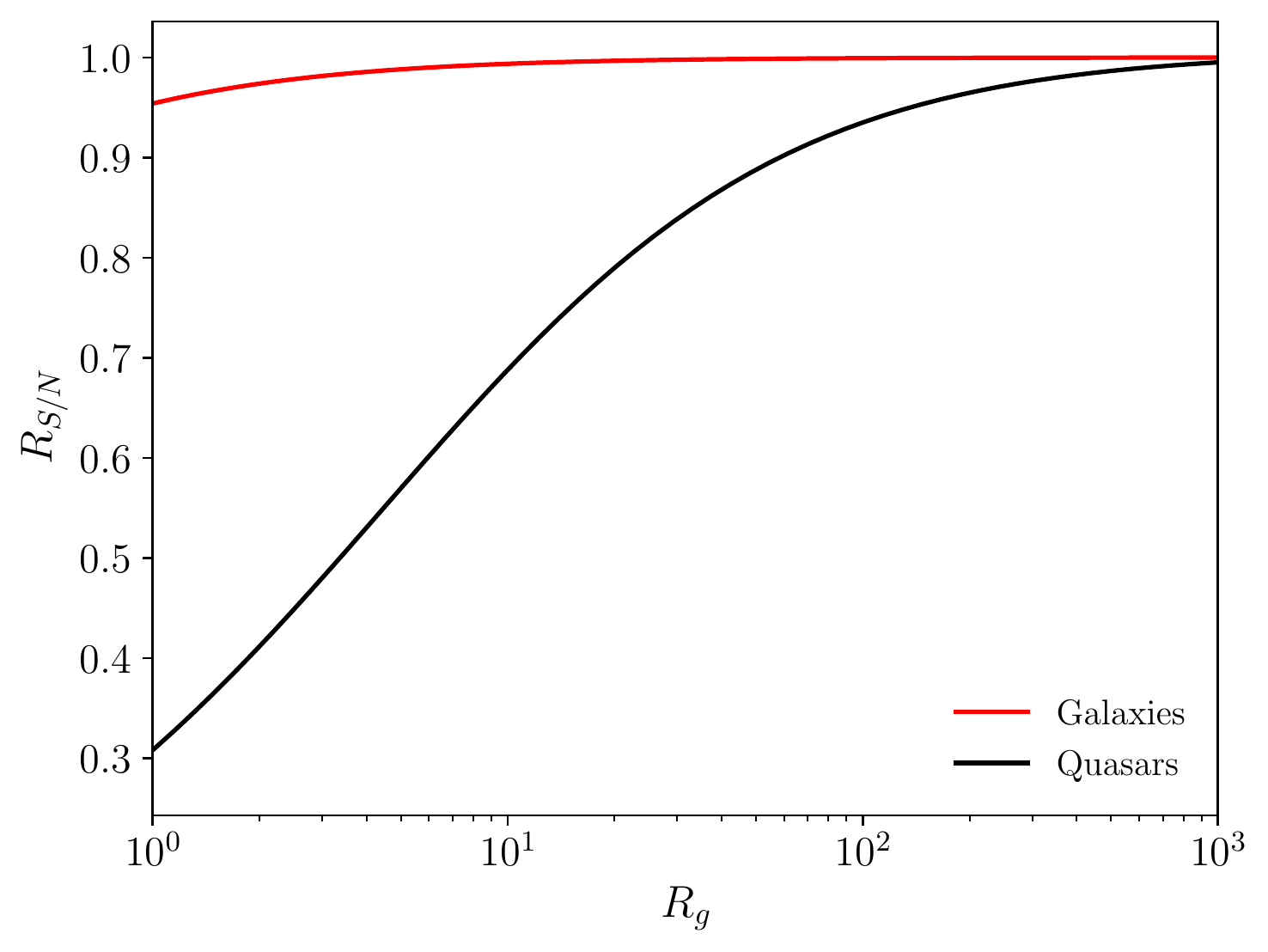}
      \caption{Relative improvement in the $S/N$ of the proper motion signal when cross-correlating with a galaxy sample containing a factor $R_g$ more sources than the \gaia galaxy and quasar samples (shown in red and black respectively).}
    \label{fig:SN_Ngal}
    \end{figure}
    So far we have assumed that the galaxy sample used to cross-correlate the proper motion maps with is cosmic-variance limited (i.e. $\bar{n}_gC_\ell^gg\gg1$ for all relevant scales). This is not an unreasonable approximation, as no measurement of the proper motions of this sample is needed, and thus any sample of galaxies covering the same redshift range can be used for this purpose.  In general, when a galaxy survey is not shot-noise limited, the sensitivity improves as a function of bias weighted number density $\bar{n}$. Figure \ref{fig:SN_Ngal} quantifies the impact of this assumption for the \gaia galaxy and quasar samples. In the figure, we calculate the $S/N$ for the proper motion cross-correlation against a catalog of galaxies following the same redshift distribution as the proper motion sample, but containing a factor $R_g$ higher $\bar{n}b$ (the noiseless limit is achieved in the limit $R_g\rightarrow\infty$). We find that the \gaia galaxy sample is already close to cosmic-variance limited on the relevant scales. This is not surprising since, as we saw in the previous section, the signal in this case is concentrated on very large angular scales.
    
    In turn, the quasar catalog, covering a much larger volume, is significantly sparser, and its sensitivity benefits from a denser sample to cross-correlate with, up to $R_g\sim100$. Although this is unrealistic, given the large volume covered by the quasar sample, it is worth noting that this significant increase in density is only necessary over the redshift range $z\lesssim0.5$, where the signal is concentrated (see the bottom right panel of Fig. \ref{fig:samples}). Only $\sim75,000$ of the \gaia quasars lie at these low redshifts, and thus, for $R_g=100$, one only requires a catalog containing $\sim7.5\times10^6$ objects below $z=0.5$. Such catalogs are already available (e.g. the WISE$\times$SuperCOSMOS photometric sample of \cite{Bilicki_2016}, or the DESI Legacy Survey \citep{DESI}, covering a smaller sky fraction). We thus conclude that the size of the galaxy sample used to cross-correlate with is not an obstacle for the detection of this signal.

    So far we have adopted a fiducial value of the galaxy bias $b_g=1$. The value of $b_g$ however depends on the sample under study. Generally denser samples tend to have lower biases, so for the high-density \gaia sample assumed here, $b_g\simeq1$ is likely reasonable. Quasars, however, populate significantly higher-mass halos, and thus have biases in the range $b_g\sim(2,\,4)$ \citep{2017JCAP...07..017L}. The $S/N$ for the cross-correlation measurement is sensitive to the galaxy bias, especially in the shot-noise-dominated regime (see Eq. \ref{eq:sn_x}), and thus this assumption could affect the forecast $S/N$.

    \begin{figure}
        \centering
        \includegraphics[width=0.49\textwidth]{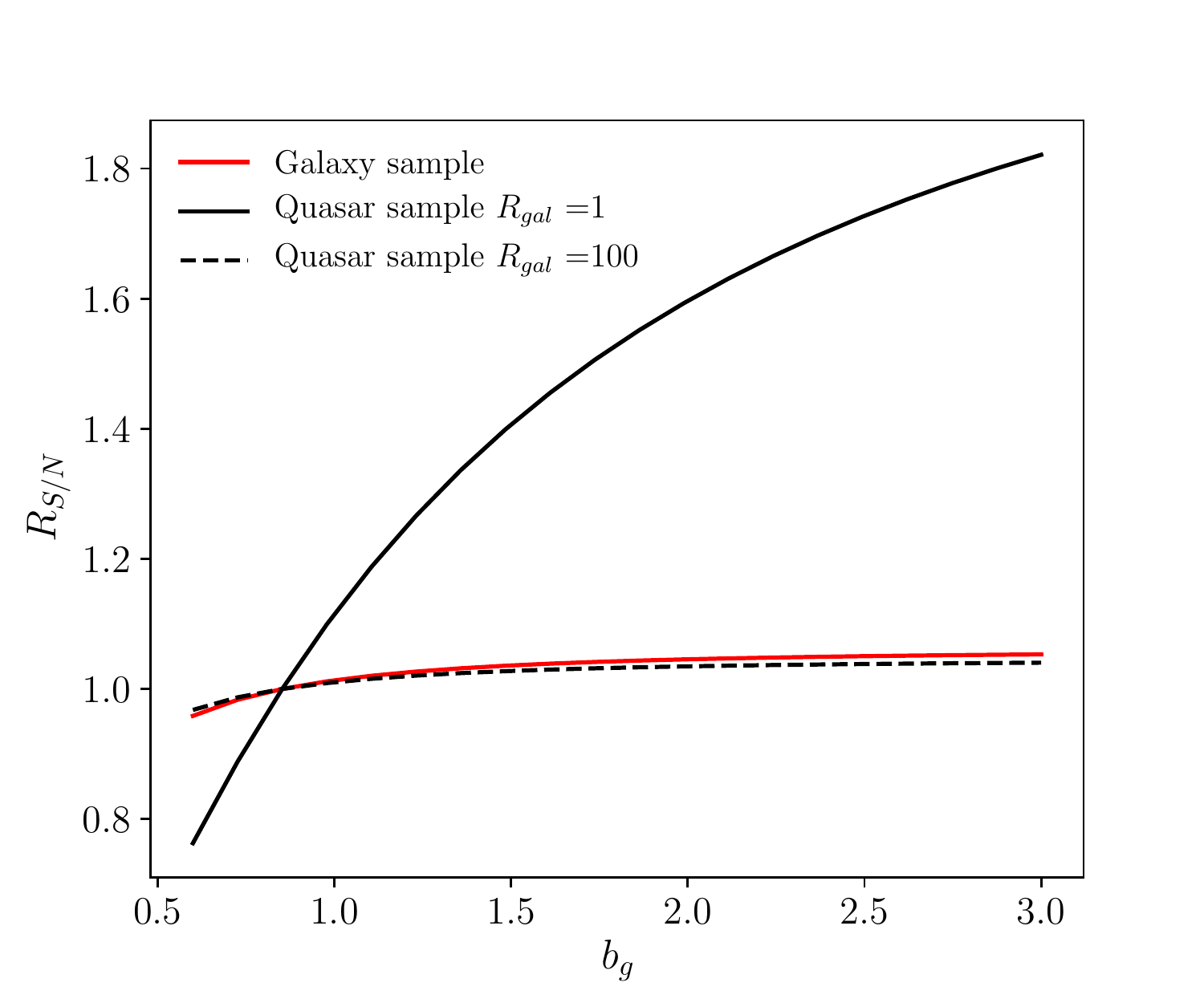}
        \caption{Fractional impact on $S/N$ of varying the galaxy bias from the fiducial value of $b_g = 1$, for the \gaia galaxy and quasar samples (shown in red and black respectively).}
        \label{fig:SN_bg}
    \end{figure}
    In Figure \ref{fig:SN_bg}, we calculate the $S/N$ for for the proper motion cross-correlation against a catalog of galaxies following the same redshift distribution as the proper motion sample, with a bias parameter $0.6 < b_g < 3$. This process was repeated for the \gaia galaxy sample and the quasar sample with and without shot-noise. The values shown in the figure correspond to the fractional improvement on $S/N$ for the case with $b_g=1$, for each of these samples. We find that the results for the quasar sample vary strongly with $b_g$, when it is not cross-correlated by a higher-density sample, reaching $1.8\times S/N$ at $b_g = 3$. However, when the quasar sample is cross-correlated with a large enough sample (for which, as we mentioned, data already exists), the dependence on $b_g$ is significantly smaller, varying by $5\%$ of the fiducial $S/N$ across the range of $b_g$. This variation is comparable to the dependence of the galaxy measurement on $b_g$. We therefore expect the results presented in the previous sections to be largely insensitive to our choice of $b_g$.

\section{Conclusions}\label{sec:conc}
  In this paper we have studied detectability of cosmic proper motions by \gaia as an example of the current generation of proper-motion surveys. The main observable is the time variation in the angular position of extragalactic sources caused by their peculiar velocities, and therefore sensitive to the growth of structure. We find, that the best approach towards detecting the cosmic proper motions is though cross-correlation with galaxy surveys, which are, especially at low redshift, mostly already nearly sample variance limited.  

  We find that, against a na\"ive expectation, based on the amplitude of this signal for individual objects, the cumulative signal-to-noise with \gaia is not too far from unity, with the galaxy and quasar samples achieving $S/N\sim0.3$ and $\sim0.15$ respectively. Galaxies have noisier proper motion measurements, but they have considerably higher density at low redshifts, where the proper motions are larger for the same velocity field.  The results can be brought above $S/N=1$ by either having more galaxies or integrating for longer. We reiterate that proper motions, like measuring the real-time expansion of the universe, have a very favourable error scaling with observation time of $T^{3/2}$, since the total motion of an object on the sky grows linearly with time, while the error drops as inverse root of time. We find that a survey that has either $\sim$10 times as many sources and integrates $\sim$2.5 times longer, or that simply has $\sim$15 times the precision of \gaia, could reach a detection of cosmological peculiar motions with $S/N\sim5$. This can be compared with the factor $\sim O(100)$ improvement in the astrometric precision of \gaia with respect to its precursor Hipparcos \citep{1997A&A...323L..49P}. In other words, next generation proper motion surveys might be in position to achieve a detection. We caution that this assumes a full-sky survey and since the signal is concentrated at low angular multipoles, a survey targeting a smaller sky area, such as the Roman Space Telescope \citep{2022MNRAS.512.5311W}, would not be appropriate.

  While we find that the low-redshift galaxies are the most promising tracer, their signal-to-noise is concentrated on larger angular scales, which are more prone to systematic errors. Therefore, quasars have their own advantages in terms of getting the necessary signal-to-noise. In all cases, we find that the shot noise of the galaxy sample against which the proper motion signal is cross-correlated, does not significantly affect its detectability, and full-sky samples with the required number density and survey depth are already available.

  It must be noted, that the calibration of the proper motion measurements is based on assuming distant quasars are at rest with respect to the cosmological rest frame. This type of calibration method would essentially null out any signal coming from the lowest multipoles, with the detailed mode loss depending on the precise calibration scheme. We find that this will likely cause a significant, but not catastrophic, loss of signal to noise by a factor of around 2 (albeit with large uncertainties), unless alternative calibration strategies can be devised.

  Throughout this work we have assumed that all samples involved have perfect redshift estimates. As we discussed in Section \ref{sec:theory}, the amplitude of the proper motion signal is degraded if projected over a broad redshift range, and thus redshift errors could hamper a potential future detection. Additionally, and for this reason, any attempt at detecting this signal would likely benefit from a 3D treatment of the proper motion signal, as opposed to the multi-bin ``tomographic'' approach described here. Further insight may be gained from future releases of the \gaia data, particularly if full spectral information is available for extragalactic sources.

  In addition to the basic confirmation of our cosmic paradigm by observing the universe evolve in real time, cosmic proper motions would be one of the few dimensional cosmological numbers that we measure. This would directly tie the Hubble expansion rates to our clocks, rather than indirectly via other dimensional measurements (such as CMB temperature and knowledge of fundamental physical constants). Therefore it offers one of the few ways of clarifying the Hubble tension, assuming it still exists when such measurements are finally within instrumental sensitivity. In details, we are mostly sensitive to the product of $H f b\sigma_8$, where $b$ is the galaxy bias. The product of $f b \sigma_8$ can be measured for the given galaxy sample by clustering and redshift-space distortions allowing one to measure $H_0$ directly.

\begin{acknowledgments}
We would like to thank David Hogg for useful discussions, and the anonymous referee for their insightful comments, which improved the quality of this paper. DA acknowledges support from the Beecroft Trust, and the Science and Technology Facilities Council through an Ernest Rutherford Fellowship, grant reference ST/P004474.  K.S.F. is supported by the NASA FINESST program under award number 80NSSC20K1545.
\end{acknowledgments}

\bibliography{main}

\appendix
\section{Lensing drift}\label{app:lensing}
  Gravitational lensing causes a displacement in the angular position of a source at comoving distance $\chi_s$, given by
  \begin{equation}
    \boldsymbol{\alpha}_L=\nabla_{\nv}\int_0^{\chi_s}d\chi\frac{\chi_s-\chi}{\chi_s\chi}\,[\psi+\phi](\chi\nv,\eta=\eta_0-\chi),
  \end{equation}
  where $\psi$ and $\phi$ are the Newtonian-gauge metric potentials, $\eta$ is conformal time, and $\eta_0$ its the value of $\eta$ at the observer. As the gravitational potentials decay or grow, gravitational lensing causes a slow drift in $\boldsymbol{\alpha}$, which contributes to the overall apparent proper motion of sources.

  Differentiating with respect to $\eta_0$ (which coincides with the standard cosmic time $t$ at $z=0$), we obtain an expression for this lensing-induced drift:
  \begin{equation}
    \tpm_L=\nabla_{\nv}\int_0^{\chi_s}d\chi\frac{\chi_s-\chi}{\chi_s\chi}\,\partial_\eta[\psi+\phi].
  \end{equation}
  In $\Lambda$CDM, $\psi=\phi$, and the Newtonian potential is related to the matter overdensity via the Poisson equation which, in Fourier space, reads
  \begin{equation}
    \psi({\bf k})=-\frac{3}{2}H_0^2\Omega_M\frac{\delta({\bf k})}{ak^2}.
  \end{equation}
  Assuming self-similar linear growth for $\delta$ (valid in the linear regime), the time derivative of $\psi+\phi$ is then given by
  \begin{equation}
    \partial_\eta(\phi+\psi)=aH(f-1)\,(\phi+\psi).
  \end{equation}

  Following the same steps outlined in Section \ref{ssec:theory.cl}, combined with the results above, the transfer function for the lensing drift $E$-mode is given by
  \begin{equation}
    Q_\ell^{L}(k,\chi)=\frac{\sqrt{\ell(\ell+1)}}{(\ell+1/2)^2}\frac{2H(z)\,[1-f(z)]}{1+z}q_L(\chi)j_\ell(k\chi),
  \end{equation}
  where $z$ is the redshift to comoving distance $\chi$, and the lensing kernel $q_L$ is
  \begin{equation}
    q_L(\chi)=\frac{3H_0^2\Omega_M}{2}(1+z)\chi\,\int_{z(\chi)}^\infty dz'\,p(z')\left(1-\frac{\chi}{\chi(z')}\right).
  \end{equation}

\begin{figure}
    \centering
    \includegraphics[width=0.6\textwidth]{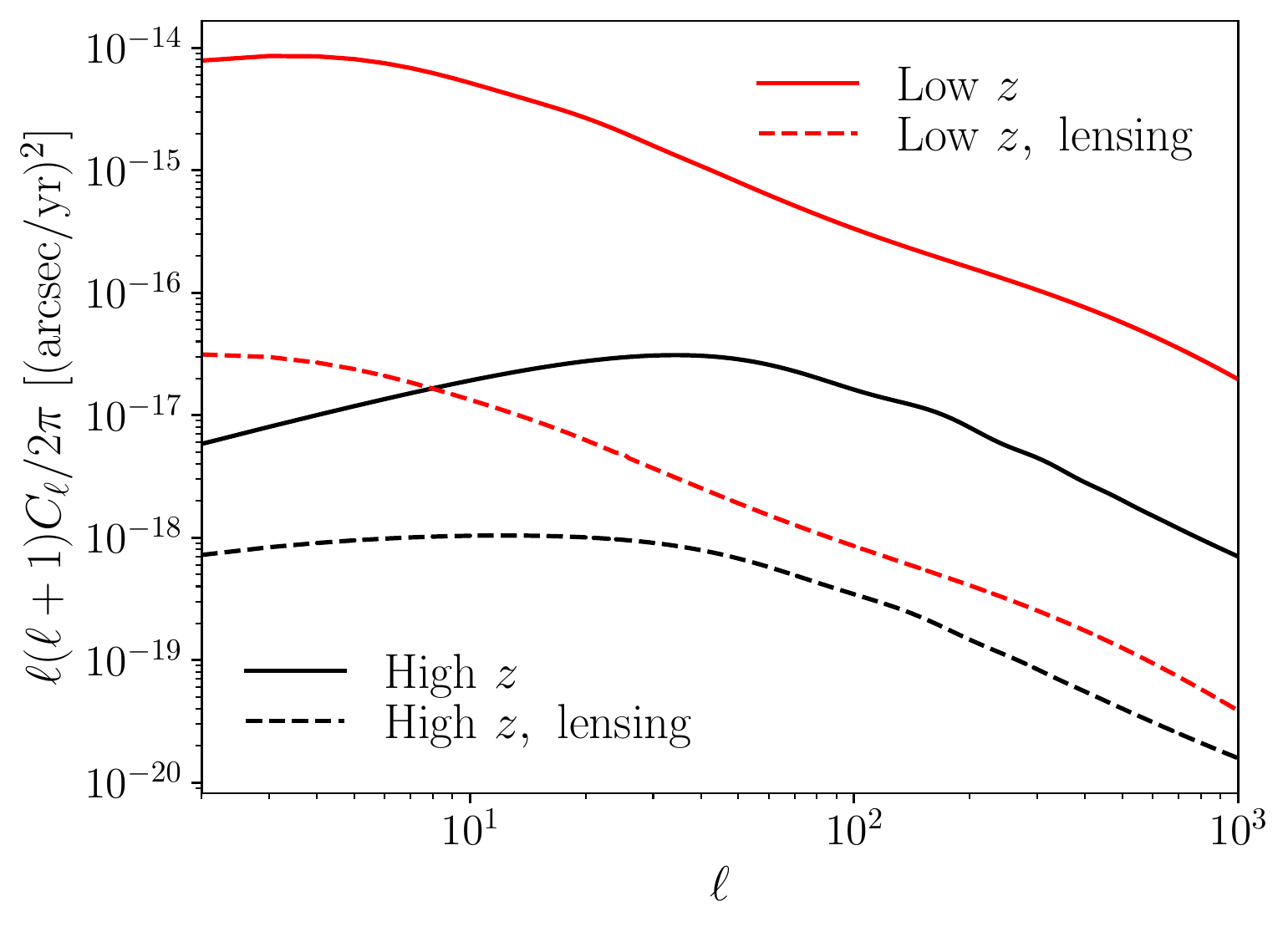}
    \caption{Angular power spectrum of the proper motion signal for a sample of galaxies at low redshifts ($z\sim0.1$, red), and at high redshifts ($z\sim1$, black). The total power spectrum, including the intrinsic signal and the lensing drift contribution, is shown as solid lines, whereas the lensing drift component is shown as dashed lines. The latter is subdominant in all cases, and hence likely undetectable.}
    \label{fig:lensdrift}
  \end{figure}

  Two important conclusions can be gleaned from these results.
  \begin{itemize}
    \item First, unlike the contribution from lensing magnification to the observed galaxy overdensity, the lensing drift effect has the same sign as the intrinsic proper motion. This is because, although lensing tends to displace observed positions {\sl away} from a foreground overdensity, the decaying nature of the gravitational potential at late times (i.e. $f<1$) causes the lensing contribution to the apparent proper motion to point {\sl towards} the overdensity.
    \item Second, although lensing is a cumulative effect, and thus source displacements are larger for higher-redshift sources, the lensing drift effect is sourced by time-varying gravitational potentials, and thus only kicks in at late times, after the matter-dominated era (e.g. as in the case of the integrated Sachs-Wolfe effect). The effect thus saturates at $z\sim1$.
  \end{itemize}
  To estimate the lensing drift contribution to the proper motion signal we consider two configurations: a low-redshift sample centered at $z=0.1$, with a Gaussian redshift distribution of width $\sigma_z=0.05$, and a high-redshift sample at $z=1$ with a width $\sigma_z=0.2$. Figure \ref{fig:lensdrift} shows the angular power spectrum of the proper motion $E$-mode including both the intrinsic signal and the lensing drift contribution, in solid black and red for the low-redshift and high-redshift samples respectively. The contribution from lensing (including both the lensing drift auto-correlation and its cross-correlation with the intrinsic signal) is shown as dashed lines for both cases. The lensing contribution is always subdominant and likely undetectable. Quantitatively, at low redshifts the standard deviation of the intrinsic peculiar motion signal is $\sim0.14\,\uasy$, while the lensing contribution is roughly three orders of magnitude smaller. At higher redshifts, the lensing contribution can be up to $\sim15\%$ of the total signal, although the intrinsic peculiar motion signal itself is more than 10 times smaller than at low redshifts.
\end{document}